\documentclass[aip,amsmath,amssymb,preprint]{revtex4-1}

\usepackage{graphicx}
\usepackage{dcolumn}
\usepackage{bm}

\usepackage[utf8]{inputenc}
\usepackage[T1]{fontenc}
\usepackage{mathptmx}
\usepackage{etoolbox}

\makeatletter
\def\@email#1#2{%
 \endgroup
 \patchcmd{\titleblock@produce}
  {\frontmatter@RRAPformat}
  {\frontmatter@RRAPformat{\produce@RRAP{*#1\href{mailto:#2}{#2}}}\frontmatter@RRAPformat}
  {}{}
}%
\makeatother

\draft 

\begin{document}

\preprint{AIP/123-QED}

\title[Charge to spin conversion in atomically thin bismuth]{Charge to spin conversion in atomically thin bismuth}
\author{Wilson J. Y\'{a}nez-Parre\~{n}o}
 \affiliation{Department of Physics, Pennsylvania State University, University Park, Pennsylvania 16802 USA}
\author{Alexander Vera}
\affiliation{Department of Materials Science and Engineering, The Pennsylvania State University, University Park, Pennsylvania 16802 USA}%
\affiliation{Center for Nanoscale Science, The Pennsylvania State University, University Park, Pennsylvania 16802 USA}
\author{Sandra Santhosh}
 \affiliation{Department of Physics, Pennsylvania State University, University Park, Pennsylvania 16802 USA}
\author{Chengye Dong} 
\affiliation{2-Dimensional Crystal Consortium, The Pennsylvania State University, University Park, Pennsylvania 16802 USA}
\affiliation{Center for 2-Dimensional and Layered Materials, The Pennsylvania State University, University Park, Pennsylvania 16802 USA}
\author{Jimmy C. Kotsakidis}
\affiliation{Laboratory for Physical Sciences, College Park, Maryland 20740 USA}
\author{Yongxi Ou}
\author{Saurav Islam}
 \affiliation{Department of Physics, Pennsylvania State University, University Park, Pennsylvania 16802 USA}
\author{Adam L. Friedman}
\affiliation{Laboratory for Physical Sciences, College Park, Maryland 20740 USA}
\author{Maxwell Wetherington}
\affiliation{Materials Research Institute, The Pennsylvania State University, University Park, Pennsylvania 16802 USA}
\author{Joshua Robinson}
\affiliation{Department of Physics, Pennsylvania State University, University Park, Pennsylvania 16802 USA}
\affiliation{Department of Materials Science and Engineering, The Pennsylvania State University, University Park, Pennsylvania 16802 USA}%
\affiliation{Center for Nanoscale Science, The Pennsylvania State University, University Park, Pennsylvania 16802 USA}
\author{Nitin Samarth}
\email{nsamarth@psu.edu}
\affiliation{Department of Physics, Pennsylvania State University, University Park, Pennsylvania 16802 USA}
\affiliation{Department of Materials Science and Engineering, The Pennsylvania State University, University Park, Pennsylvania 16802 USA}
\affiliation{Center for Nanoscale Science, The Pennsylvania State University, University Park, Pennsylvania 16802 USA}%

\date{\today}

\begin{abstract}
We report charge to spin conversion in a hybrid heterostructure comprised of atomically thin bismuth (Bi) confined between a silicon carbide (SiC) substrate and epitaxial graphene (EG). We confirm composition, dimensionality, and a 96.5 \% intercalation coverage using X-ray photolectron spectroscopy, scanning transmission microscopy, low energy electron diffraction, and Raman spectroscopy. Electrical transport measurements show signs of weak antilocalization in the heterostructure, consistent with spin-orbit coupling in this hybrid heterostructure. Spin torque ferromagnetic resonance measurements in permalloy/EG/2D-Bi heterostructures probe charge-to-spin conversion and revealing that an in plane polarization of the spin current, perpendicular to the charge current. The ratio of the in-plane to out-of-plane torque is 3.75 times higher than in hydrogenated graphene control samples.
\end{abstract}

\pacs{}

\maketitle 

\section{Introduction}

Electrically driven charge-spin interconversion is a phenomenon of great interest in contemporary spintronics because it provides an attractive route toward energy-efficient, high density, non-volatile magnetic random access memory. \cite{Hellman_RevModPhys.89.025006,Fert_RevModPhys.96.015005} Bismuth (Bi) is an important material in this context because of the large strength of the spin-orbit coupling (SOC) which makes it attractive for spin-orbit torque (SOT) devices that rely on spin current generated by an electrical current. Despite this naive expectation of efficient SOT based on the strong SOC in Bi, the published reports on charge-spin conversion efficiency in Bi show large discrepancies (several orders of magnitude) in the key figure of merit, the spin Hall angle.  \cite{Hou_APL_2012,Emoto_PhysRevB.93.174428,Yue_PhysRevLett.121.037201,Sanchez_NatComm,ou2023spinhallconductivitybi1xsbx}. It is suspected that the ordinary Nernst effect may have played a role in these discrepancies.\cite{Liang_PhysRevB.106.L201304} Additionally, it is now recognized that the strongly anisotropic Fermi surface of Bi can lead to significant variations in the spin-charge interconversion with crystalline orientation. \cite{Liang_PhysRevB.106.L201304,Fukumoto_PNAS_Bi,corbae2024} These results, as well as the continued interest in spin-charge conversion in Bi from a fundamental theoretical perspective, \cite{Qu_PhysRevB.107.214421} motivate us to explore spin transport in Bi at another frontier, namely in the true two-dimensional (2D) limit where 2D-Bi has attracted significant attention for realizing topological phenomena such as the quantum spin Hall insulator. \cite{Murakami_PhysRevLett.97.236805,Wada_PhysRevB.83.121310,Reis_Science} Nevertheless, ultra-thin synthesis of Bi on an insulating substrate in a manner that is resilient to oxidation is extremely challenging, preventing explorations ofthe 2D regime of Bi films within the context of spintronics. To this end, we report the synthesis of air-stable, wafer-scale, crystalline, 2D-Bi films on an insulating substrate in a manner that naturally protects against oxidation. We then investigate spin-charge conversion in these films using spin torque ferromagnetic resonance (ST-FMR) measurements. 

\section{Structural characterization}

We synthesize ultrathin 2D-Bi films confined between a 6H-SiC substrate and epitaxial graphene via confinement heteroepitaxy (CHet, see Methods).\cite{Briggs2020} In this approach, Bi atoms sublimate from a metallic Bi precursor at elevated temperatures and near atmospheric pressures, then diffuse through a defective EC interface on SiC to form an ultra-thin, 2D-Bi layer sandwiched between EG and SiC. To assess the efficacy of intercalation and the resultant structure, we observe changes in characteristic features of EG on SiC in X-ray photoelectron spectroscopy (XPS) and low electron energy diffraction (LEED), as performed in numerous previous reports. \cite{Riedl_2010,Forti_2014,C9NR03721G,WU2021100637} The C $1s$ spectrum, shown in Fig. 1(a), hosts a largely asymmetric peak at 284.8 eV and a slightly weaker and more symmetric peak at 283.1 eV, ascribed to the graphene and SiC, respectively. Intercalation is evidenced by elimination of the buffer layer peaks between 285 – 286 eV (which are absent within our detection limit here) and a characteristic ``split'' in the C $1s$ spectra associated with the redshift of the SiC peak from its native position ($\sim$283.8 eV). \cite{PhysRevLett.103.246804} This shift is presumed to be related to band bending of SiC, \cite{PhysRevLett.108.246104} as the replacement of the buffer layer with Bi atoms modifies the surface dipole and thus core energy levels for carbon atoms in SiC near the surface. The Bi 4f spectrum in Fig. 1(b) has the previously absent spin split Bi $4f_{7/2}$ (157.2 eV) and Bi $4f_{5/2}$ (162.5 eV) peaks arising due to the Bi intercalant. The position and asymmetric shape of these peaks evidences the metallic nature of 2D-Bi, even after being exposed to ambient conditions after growth. The evolution of the C $1s$ and Si $2p$ spectra can be seen in the supplementary material.

An additional characteristic feature of intercalation between the buffer layer and SiC is the reduction of the ($6\sqrt{3}\times6\sqrt{3}$)R30$^\circ$ superstructure spots in LEED around both the SiC$\{01\}$ and graphene$\{01\}$ spots, and a heightened intensity ratio of graphene$\{01\}$ to SiC$\{01\}$.\cite{PhysRevLett.103.246804} Within the resolution limit, the LEED image (Fig. 1(c)), taken at 110 eV incident beam energy, shows only the first order diffraction spots of graphene and SiC and no ($6\sqrt{3}\times6\sqrt{3}$)R30$^\circ$ reconstruction, implying the buffer layer has been eliminated by Bi intercalant atoms. Notably, no additional superstructure spots emerge; this result diverges from previous reports on ion bombardment and deposition-based intercalation of Bi which detail the emergence of SiC($\sqrt{3}\times\sqrt{3}$)R30°  \cite{PhysRevB.94.085431,Sohn2021} or an enhancement of the SiC$(6 \times 6)$ spots due to Bi’s epitaxial registry to SiC. \cite{Sohn2021} However, the LEED image bears similarity to that of other CHet-based intercalant species H, Ga, In, and Sn, \cite{C9NR03721G, Briggs2020} as well as experiments on Dy, \cite{PhysRevB.107.045408} Sb, \cite{https://doi.org/10.1002/andp.201900199} Ag,\cite{Rosenzweig_PhysRevB.101.201407} and Au  intercalation\cite{Forti_2020} which commonly ascribe this pattern to either a SiC$(1 \times 1)$ phase or, in the case of Sb and Dy, a lack of long-range order in the intercalant layer; hence, both cases may be possible here.

To better assess the nanostructure of intercalated Bi, a cross-sectional slab ($\sim 20 \mu$m) extracted from the sample via focused ion beam (FIB) milling is examined under annular dark field scanning transmission electron microscopy (ADF-STEM) and energy dispersive spectroscopy (EDX) \cite{doi:10.1021/acsami.1c14091} along the [1120] zone axis of SiC, as shown in Fig. 1(d). A high magnification window shows a locally ordered bilayer of Bi atoms, confirmed by the EDX signal in the inset. However, imaging conditions preclude the simultaneous observation of Bi and SiC, thus lateral epitaxy is difficult to directly resolve. Decreasing the magnification allows visualization of this bilayer Bi across larger windows, appearing as a thin, bright horizontal line at the surface of SiC (see supplemental material). The continuity of this line across the $\sim 1 \mu$m window highlights the high coverage of this predominantly bilayer Bi in the window scanned. Additional bright patches can be seen as well, which are likely BiO$_x$ or SiO$_x$ particles that remain after synthesis or are developed during the FIB process.

A larger lateral assessment of coverage may be seen with optical microscopy \cite{PhysRevMaterials.5.024006} and Raman spectroscopy mapping, as demonstrated in Fig. 1(e) and 1(f). Within the optical micrograph, beyond the bright, semi-parallel lines which range up to 10 $\mu$m apart (this contrast derives from multilayer graphene at step-edges in SiC, \cite{cryst6050053} a bright and dark region can be identified. A Raman acquisition taken within this bright region has broad, but distinct spectroscopic peaks within the ``ultra-low frequency'' (ULF) regime below the folded transverse acoustic mode of 6H-SiC (10 – 150 cm$^{-1}$) which are absent in dark regions and in EG/SiC before intercalation (see supplementary material). Similar ULF peaks have been reported in prior studies, \cite{Wetherington_2021}, highlighting a strong correlation of these features to a previous study on Bi-III under high pressure. \cite{doi:10.1021/acs.jpcc.0c08371} This suggests that these peaks are indicative of intercalated 2D-Bi. We also rule out the contribution from bismuth oxides: although bismuth oxide can exhibit features in this range, \cite{DEPABLOSRIVERA2021157245} the low O 1s signal from XPS implies minimal oxides at the surface, which would not correlate well with the lateral consistency of the ULF feature. Hence, this feature is utilized to map the surface, finding 96.5 \% of a $10 \times 10\ \mu$m map exhibits this feature and is intercalated. Lastly, we call attention to the higher frequency Raman regime, containing the D ($\sim$1350 cm$^{-1}$), G ($\sim$1596 cm$^{-1}$) and 2D ($\sim$2721 cm$^{-1}$) peaks associated with graphene. \cite{PhysRevB.99.045443} The position and full width half maximum (FWHM) of both the G and 2D peaks are consistent across the $10 \times 10\ \mu$m map outside of step-edges, indicative of a uniform, bilayer graphene along terraces (see supplemental material). A lower D/G ratio after intercalation results from catalytic ``healing'' of the previously defective graphene, like other CHet-based 2D metals. \cite {Briggs2020,Wetherington_2021}

Finally, assuming a SiC($1 \times 1$) phase is present here, we find a discrepancy with prior work in which the aforementioned SiC($6 \times 6$) enhancement in the $\alpha$-phase (also with SiC($1\times1$) epitaxy) of intercalated Bi \cite{Sohn2021} was observed under only the buffer layer. The presence of an additional attenuating graphene layer \cite{PhysRevMaterials.4.124005}, a lack of long-range order in this work, and a potential lack of lateral intercalation coverage in this previous work may all contribute to this discrepancy. It is too speculative to ascribe a structure given our experimental data; however, we recommend continued synthesis improvements (such as different temperatures, \cite{MAMIYEV2022102304} longer cool times or post-anneals, \cite{KIM2020229} and more precise defect engineering),accompanied by additional structural characterization to address this difference.



%
%
 \begin{figure}
 \includegraphics[width=90mm]{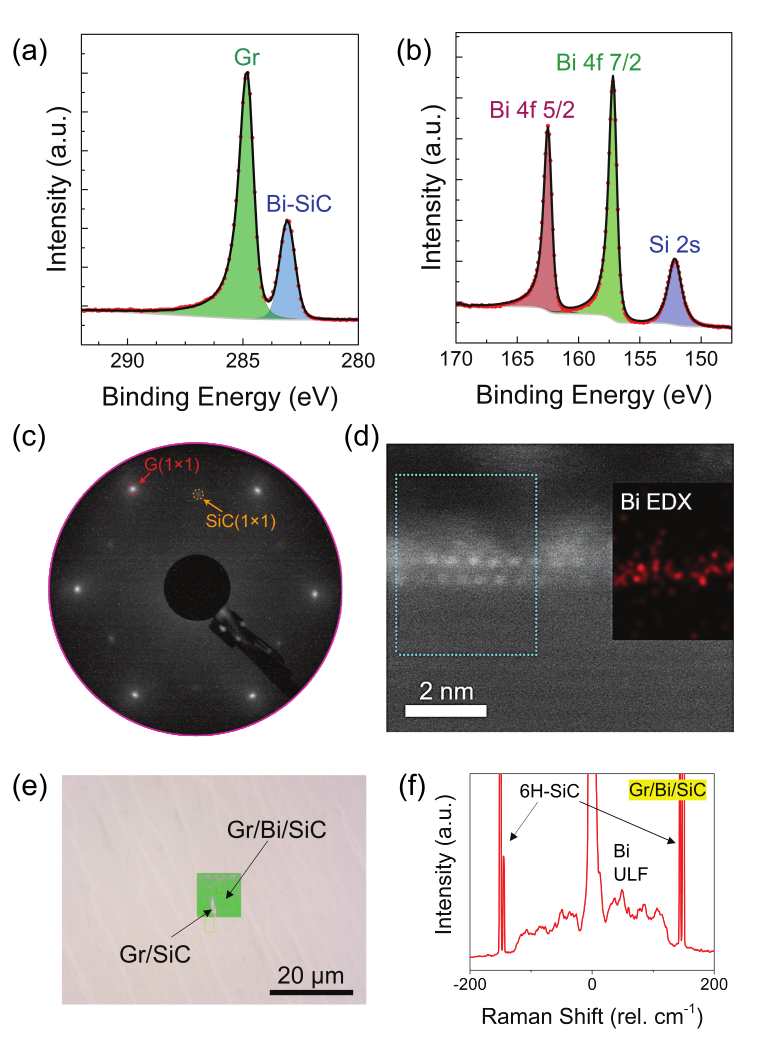}%
 \caption{\label{Fig1} (a) C $1s$ and (b) Bi $4f$ core photoelectron spectral windows taken on a Gr/Bi/SiC sample. For the C $1s$ spectral window (280 - 291.5 eV) two peaks are apparent: an asymmetric peak at 284.8 eV and a peak at 283.1 eV associated with graphene and SiC, respectively. Within the Bi $4f$ window (145 - 170 eV) three peaks are captured; two associated with metallic Bi at 162.5 eV and 157.2 eV, and a peak associated with Si 2s photoelectrons from SiC at 152.3 eV. (c) Shows a low energy electron diffraction image taken with 110 eV beam energy. Only graphene and SiC diffraction spots are visible, implying intercallation has eliminated the buffer layer. (d) A $\sim$8 nm window from high-resolution scanning transmission electron microscopy taken in the high angle annular dark field mode. A hazy bilayer of Bi atoms is seen; affirmed by electron energy loss spectroscopy map taken in the same region.    (e) Optical micrograph of a Gr/Bi/SiC sample, overlaid with a map of the intensity of the ultra-low frequency Raman feature seen in (f). A large percentage ($96.5\%$) of the sample surface in the scanned region exhibits this feature, suggesting micron-scale coverage of Bi intercalant.}%
 \end{figure}

\section{Electrical transport characterization}

To characterize the electrical transport properties of the EG/Bi/SiC heterostructure, we fabricated Hall bars using standard photolithography and Ar plasma etching (Fig. 2 (a)). The dimensions of the Hall bar devices (length, width) are 100 $\mu$m $\times$ 50 $\mu$m, respectively. The samples show insulating behavior with an increase in resistance as temperature is reduced. The symmetrized longitudinal sheet resistance ($R_{xx}$) and anti-symmetrized Hall resistance ($R_{yx}$) of the sample as a function of magnetic field is shown in Fig. 2(c) and 2(d). Typical carrier density and mobility of the samples extracted from the Hall coefficient and sheet resistance at $T=2$~K are $\approx10^{13}$~cm$^{-2}$ and $330$~cm$^2$/Vs with electrons being the dominant carrier type. Note that these numbers represent the combined charge transport through two parallel channels, EG and 2D-Bi.  

At $T=300$~K, $R_{xx}$ shows parabolic dependence on the magnetic field, characteristic of the expected classical magnetoresistance. At $T=2$~K, we observe positive magnetoresistance with a magnetic field dependence characteristic of weak antilocalization (WAL) consistent with the presence of strong SOC, as seen in many materials including topological insulators and epitaxially grown Bi. \cite{PhysRevLett.98.136801,PhysRevB.104.075431,PhysRevB.83.245438}. We note that even though WAL can also arise in EG in the absence of SOC,\cite{PhysRevB.99.245407} the EG/SiC template samples used for CHet always show weak localization (WL). Strikingly, EG/2D-Pb/SiC CHet samples also show negative magnetoresistance at low temperature, consistent with WL despite the expected presence of strong SOC in Pb (See supplementary information) \cite{https://doi.org/10.48550/arxiv.2205.06859}. This suggests that there might be other factors (beyond SOC) at play in CHet systems. Nevertheless, our observation of positive magnetoresistance in EG/2D-Bi/SiC heterostructures is a clear indication that Bi contributes strongly to the electronic transport. The presence of two parallel conducting channels (EG and 2D-Bi), however, presents a significant obstacle to detailed analysis of the magnetotransport in the heterostructures.

 \begin{figure}
 \includegraphics[width=130mm]{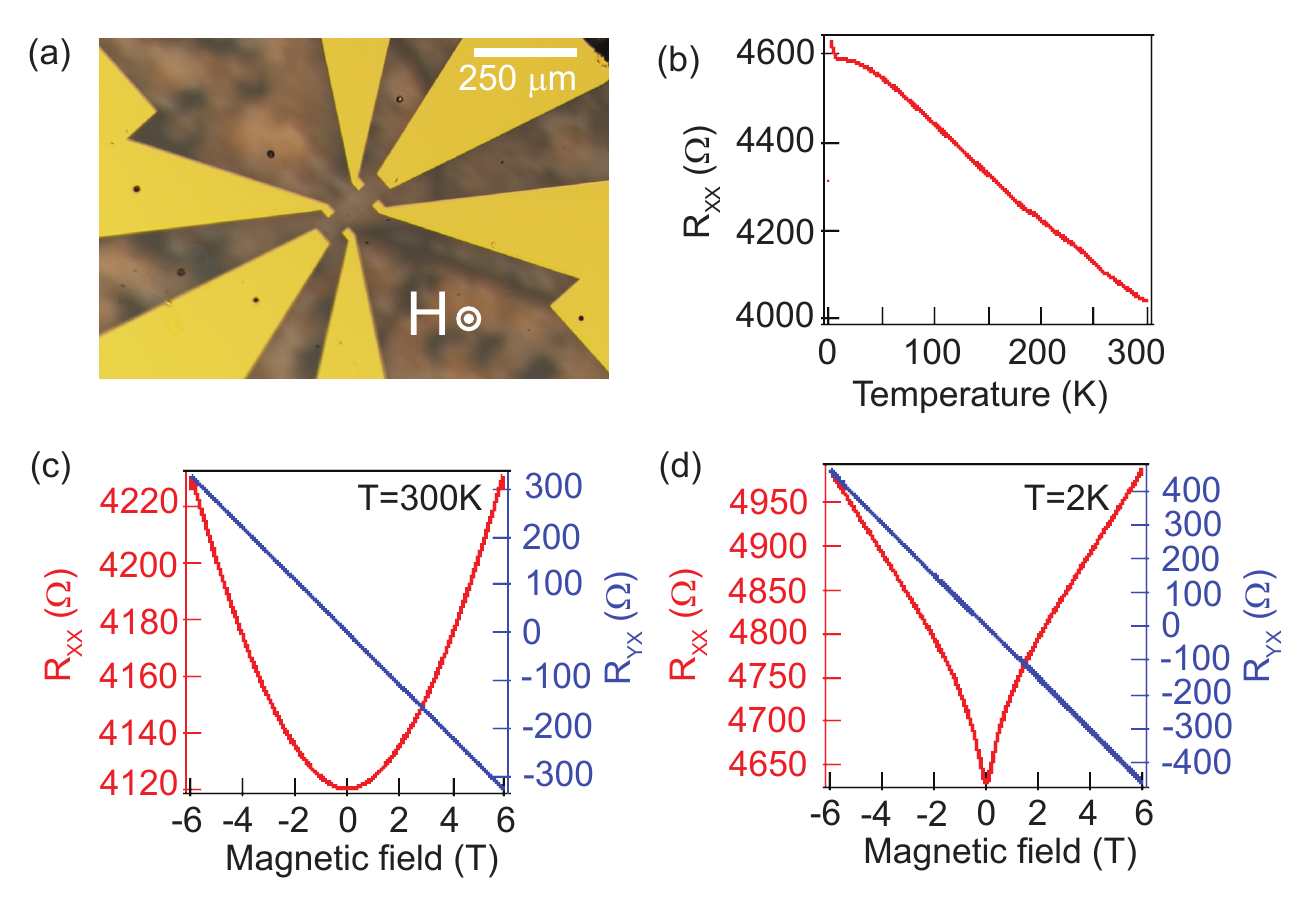}%
 \caption{\label{Fig2} (a) Optical image of a Hall bar used for electrical transport characterization. (b) Resistance of a Bi/EG sample versus temperature. Longitudinal ($R_{xx}$) and transverse ($R_{yx}$) resistance as a function of magnetic field at (c) T = 300 K and (d) T= 2 K in Bi/EG. }%
 \end{figure}

 \section{Charge to spin conversion in bismuth}
We now discuss the charge-to-spin conversion process in 2D-Bi/EG heterostructures that are interfaced with a conventional ferromagnet, Ni$_{0.8}$Fe$_{0.2}$ (permalloy, Py). The Py overlayer is deposited at room temperature from a single source using an electron beam evaporator with a pressure lower than $2.5 \times 10^{-7}$ torr. We deposit 4 and 8 nm of permalloy and 2 nm of aluminum to prevent oxidation of the ferromagnet. We use ST-FMR to carry out measurements of the charge-to-spin conversion process in our CHet-grown 2D-Bi films. These measurements are carried out at room temperature on devices patterned from SiC/2D-Bi/EG/Py heterostructures. We have measured a total of 8 devices (similar to the one shown in Fig. 3 (b)) with dimensions of $50 \times 10\ \mathrm{\mu}$m and $50 \times 20\ \mathrm{\mu}$m and oriented at $30^\circ$, $45^\circ$, $60^\circ$, $90^\circ$ and $135 ^\circ$ measured from the edge of the substrate. An in-plane external magnetic field ($\vec{H}$) is applied at an angle to the easy magnetization direction of the Py, aligning its magnetization into this direction. A radiofrequency (rf) current flowing in the channel sets the magnetization into precession generating a spin current in the EG/2D-Bi layer due to either the spin Hall or Rashba-Edelstein effects, thereby inducing spin accumulation at the EG/Py interface and thus driving a diffusive spin current into the Py film (Fig. 3 (a)). The SOT exerted by this spin current perturbs the precession of the Py magnetization, an effect that manifests as a DC mixing voltage related to the anisotropic magnetoresistance of the ferromagnet. By sweeping the magnitude of $\vec{H}$ through the condition for FMR at fixed rf frequency, we obtain a resonant line shape (Fig. 3(c)) from which the SOT and the charge-to-spin conversion efficiency can be determined. The present device geometry does not allow us to quantitatively determine the absolute magnitude of the SOT and charge-to-spin conversion efficiency because the impedance of the devices cannot be precisely determined due to parallel conduction in the EG layer. However, we can extract a ratio of the effective fields $H_{DL}$ and $H_{FL}$ that generate the SOT (or damping-like torque and the field-like torque, respectively), which is qualitatively related to the charge-to-spin conversion efficiency shown in Fig. 3 (e). 

 \begin{equation}
     \frac{H_{DL}}{{H_{Oe}+H_{FL}}}=\frac{\tau_{\parallel}}{\tau_{\bot}}=\sqrt{1+\frac{4\pi M_{\mathrm{Eff}}}{H_{\mathrm{Res}}}}\frac{V_S}{V_A}
 \end{equation}
 
These fields together with the Oersted field ($H_{\mathrm{Oe}}$) produced by the flow of current are equal to the ratio of in plane ($\tau_{\parallel}$) and out of plane ($\tau_{\bot}$) torque in the system. This can be computed by fitting the spectra using a symmetric and antisymmetric Lorentzian distribution (Fig. 3 (d)) and taking the ratio of their amplitudes ($V_S$ and $V_A$), together with some physical parameters of the system like the resonance field ($H_{\mathrm{Res}}$) and the demagnetization field ($4\pi M_{\mathrm{Eff}})$ obtained by fitting the Kittel equation to the FMR phenomena. We find some variation in the field ratio among different samples (Fig. 3 (e) and supplemental material). On average, in six out of eight devices, the torque ratio of 2D Bi is $\frac{H_{DL}}{H_{Oe}+H_{FL}}=2.01\pm0.66$, while the other two (outliers) had a significantly larger torque ratio of $\frac{H_{DL}}{H_{Oe}+H_{FL}}=3.64\pm0.39$. We attribute this variation to inhomogeneity in Bi intercalation that effectively produces devices with different amounts of Bi that can contribute to the spin to charge conversion phenomena. We tried to search for systematic behavior in the samples that showed an enhanced torque ratio and found that both of them were $\mathrm{10 \mu m \times 50 \mu m}$ bars aligned to the SiC substrate edge with larger gold contact pads in comparison with other devices. Nevertheless, these geometric aspects should not affect the charge to spin conversion efficiency measured in our experiment.  
Even though homogeneity seems to be a severe limitation in our devices, these values are still 3.75 times the value obtained in hydrogenated EG control samples which had a torque ratio of $\frac{H_{DL}}{H_{Oe}+H_{FL}}=0.97\pm0.19$. \cite{https://doi.org/10.48550/arxiv.2205.06859} This is a clear indication that Bi is contributing to the charge to spin conversion in the heterostructure. We also note that the measured ST-FMR signal becomes cleaner when the device is aligned along the step edges of the SiC substrate and becomes noisy (almost undetectable) when the device is perpendicular to these edges. This is probably due to spatial differences in intercalation between flat areas and edges along the SiC steps. Beyond this, we do not observe systematic differences in the torque ratio due to device orientation as long as the device is not perpendicular to the step edges (see supplemental material for particular device orientation). Finally, we measured angle-dependent ST-FMR (Fig. 3 (f)), where we changed the angle ($\phi$) between the current and the external magnetic field and extracted the magnitude of the symmetric ($V_S$) and antisymmetric ($V_A$) components of the ST-FMR signal; this follows the expected symmetry and functional form ($V_{\mathrm{mix}}=\cos(\varphi) \sin(2\varphi)$) for a heavy metal with the spin in-plane and perpendicular to the electrical current. 


 \begin{figure}
 \includegraphics[width=82mm]{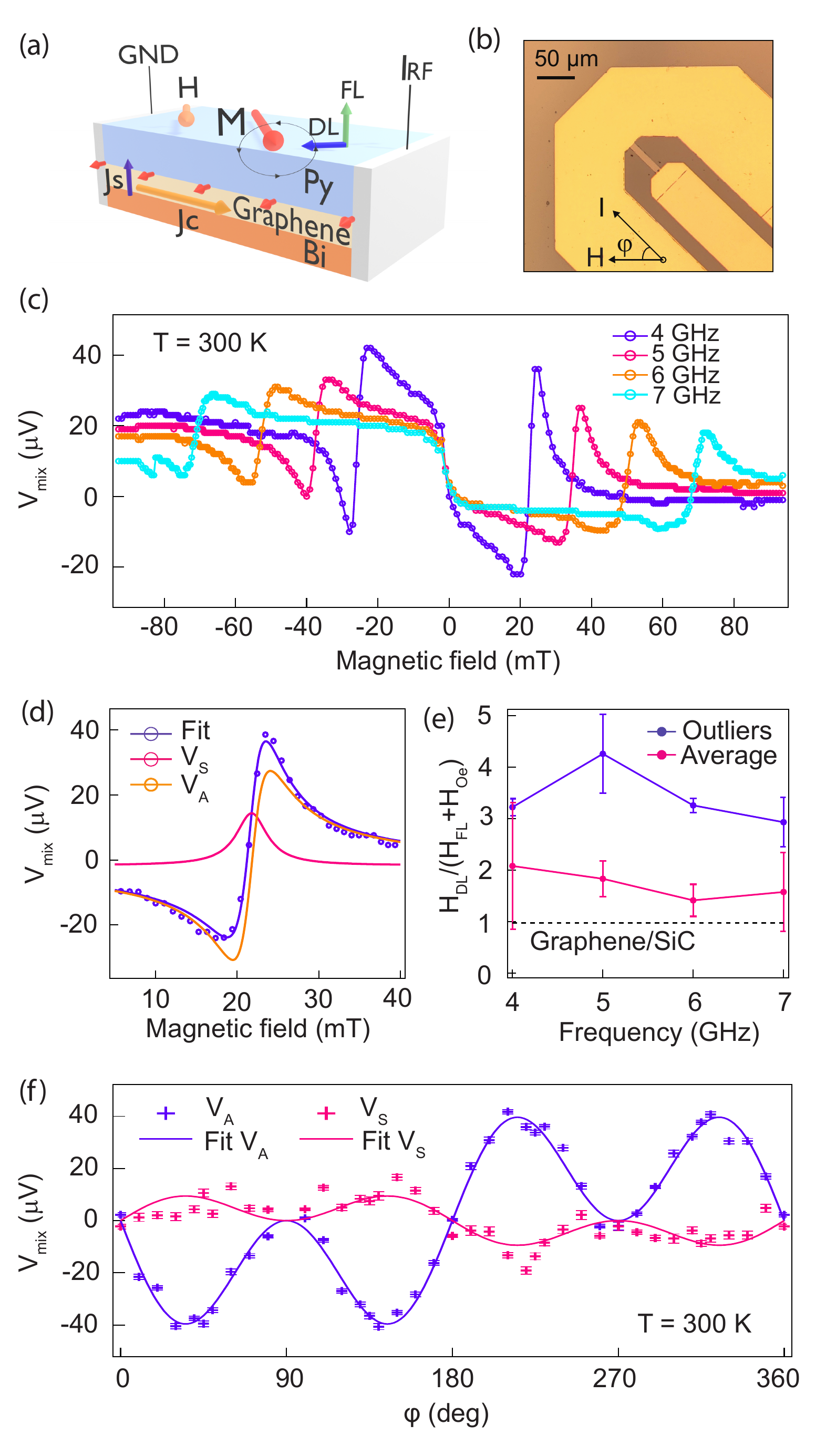}%
 \caption{\label{Fig3} (a) Schematic of the charge-to-spin conversion process in  Bi/EG/Permalloy heterostructures. (b) Optical image of a device used for ST FMR measurements. (c) Room temperature ST-FMR spectra of a Bi/EG/Permalloy (8 nm) heterostructure measured at 20 dBm, using different excitation frequencies. The magnetic field was applied at $\varphi=45^\circ$ from the current axis. (d) Fitted spectra using the 4 GHz data shown in (c). These show the dampinglike symmetric ($V_{S}$) and fieldlike antisymmetric ($V_{A}$) Lorentzian contributions used in the fit. (e)  Normalized magnitude of the measured effective fields obtained from the fit of the data shown in (c) in graphene/Bi and its comparison with a graphene control sample. We show the average torque ratio of six devices that showed similar values and the average value of two outlier devices that showed a higher torque ratio. (f) magnitude of the symmetric ($V_S$) and antisymmetric ($V_A$)
components of the mixing voltage signal obtained by changing the angle ($\varphi$) between the current and the magnetic field measured at 3 GHz and 20 dBm. These values have been fit using $V_{mix}=cos(\varphi) sin(2\varphi)$.} %
 \end{figure}

\section{Conclusions}

In summary, we have used CHet to synthesize wafer scale, air stable 2D-Bi films confined between EG and a SiC substrate. The samples are thoroughly characterized using XPS, HAADF TEM, EDX, and Raman microscopy, revealing a locally ordered bilayer of Bi with 96.5 \% intercalation. Measurements of electrical magnetotransport reveal the presence of strong SOC in these hybrid 2D-Bi/EG heterostructures via signatures of WAL, contrasting with the WL seen in CHet-grown 2D-Pb/EG samples reported in previous studies (supplementary information) \cite{https://doi.org/10.48550/arxiv.2205.06859}. ST-FMR measurements reveal charge-to-spin conversion at room temperature in Py/EG/2D-Bi heterostructures with a spin torque ratio 3.75 times higher than in hydrogenated EG control samples, albeit with significant device-to-device variation. This is probably due to inhomogeneous Bi intercalation on a macroscopic scale. Future improvements might allow us to measure similar devices in a more systematic manner. Finally, angle dependent ST-FMR measurements show that the spin polarization is in-plane and perpendicular to the flow of current. We expect that this demonstration of charge-to-spin conversion in the 2D regime serves as a proof-of-principle that spin transport can be measured in atomically thin Bi and will motivate further studies that explore the roles of SOC, dimensionality, and topology in this system.  
 


%

\begin{acknowledgments}
This work was primarily supported by the Penn State MRSEC Center for Nanoscale Science via NSF award DMR 2011839 (AV, CD, JAR, NS, SI, WJYP). We also acknowledge support from NSF award DMR 2002651 (AV, JAR) and the Penn State Two-Dimensional Crystal Consortium-Materials 
 Innovation Platform (2DCC-MIP) under NSF Grant No. DMR-2039351 (NS, SS, YO). XPS and Raman measurements were performed in the Materials Characterization Laboratory in the Materials Research Institute at Penn State University. LEED measurements were performed at the Laboratory of Physical Sciences at the University of Maryland. Electron microscopy was performed at the Canadian Centre for Electron Microscopy, a Canada Foundation for Innovation Major Science Initiatives funded facility (also supported by NSERC and other government agencies). EM work was funded by the US AFOSR Award FA9550-19-1-0239 and the NSERC (Natural Sciences and Engineering Research Council of Canada) Discovery Grant program. A portion of this research was conducted at the Center for Nanophase Materials Sciences, which is a DOE Office of Science User Facility. A. Vera is supported by the Alfred P. Sloan Foundation G‐2019-11435. J.C. Kotsakidis and A.L. Friedman acknowledge R. E. Butera for tool access.
\end{acknowledgments}

\providecommand{\noopsort}[1]{}\providecommand{\singleletter}[1]{#1}%

\end{document}